\documentclass[aps,pre,twocolumn,superscriptaddress,superscriptreference]{revtex4-1}
\usepackage{amsmath,bbold,bm,amssymb,scalerel,mathtools}
\usepackage{graphicx}
\usepackage{color}
\usepackage{enumitem}
\usepackage{algorithm,algpseudocode}
\usepackage{multirow}
\usepackage{colortbl,booktabs}
\usepackage{placeins}
\usepackage[usenames,dvipsnames]{xcolor}
\usepackage{tikz}
\usepackage{lipsum} 
\usepackage[colorlinks, linkcolor=blue!50!black, urlcolor=blue!50!black, citecolor=blue!50!black]{hyperref}
\usepackage{cleveref}
\crefname{equation}{Eq.}{Eqs.} 
\crefname{section}{Sec.}{Secs.} 
\crefname{figure}{Fig.}{Figs.}
\crefname{table}{Tab.}{Tabs.} 
\crefname{appendix}{App.}{App.}
\tikzset{%
	every neuron/.style={
		circle,
		draw,
		minimum size=0.5cm
	},
	neuron missing/.style={
		draw=none, 
		scale=1.5,
		text height=0.333cm,
		execute at begin node=\color{black}$\vdots$
	},
}

\newcommand\equalhat{%
	\let\savearraystretch\arraystretch
	\renewcommand\arraystretch{0.3}
	\begin{array}{c}
		\stretchto{
			\scalerel*[\widthof{=}]{\wedge}
			{\rule{1ex}{3ex}}%
		}{0.5ex}\\ 
		=%
	\end{array}
	\let\arraystretch\savearraystretch
}

\newcommand{\gj}[1]{\textcolor{black}{#1}}

\usepackage[normalem]{ulem}

\begin{document}

\title{Normalizing flows as an enhanced sampling method for atomistic supercooled liquids}

\author{Gerhard Jung}

\email[Corresponding author: ]{gerhard.jung.physics@gmail.com}

\affiliation{Laboratoire Charles Coulomb (L2C), Université de Montpellier, CNRS, 34095 Montpellier, France}

\affiliation{Laboratoire Interdisciplinaire de Physique (LIPhy), Université Grenoble Alpes, 38402 Saint-Martin-d'Hères, France}

\author{Giulio Biroli}

\affiliation{Laboratoire de Physique de l’Ecole Normale Supérieure, ENS, Université PSL, CNRS, Sorbonne Université, Université de Paris, 75005 Paris, France}

\author{Ludovic Berthier}

\affiliation{Laboratoire Charles Coulomb (L2C), Université de Montpellier, CNRS, 34095 Montpellier, France}

\affiliation{Gulliver, UMR CNRS 7083, ESPCI Paris, PSL Research University, 75005 Paris, France}

\date{\today}

\begin{abstract}
Normalizing flows can transform a simple prior probability distribution into a more complex target distribution. Here, we evaluate the ability and efficiency of generative machine learning methods to sample the Boltzmann distribution of an atomistic model for glass-forming liquids. This is a notoriously difficult task, as it amounts to ergodically exploring the complex free energy landscape of a disordered and frustrated many-body system. We optimize a normalizing flow model to successfully transform high-temperature configurations of a dense liquid into low-temperature ones, near the glass transition. We perform a detailed comparative analysis with established enhanced sampling techniques developed in the physics literature to assess and rank the performance of normalizing flows against state-of-the-art algorithms. We demonstrate that machine learning methods are very promising, showing a large speedup over conventional molecular dynamics. Normalizing flows show performances comparable to parallel tempering and population annealing, while still falling far behind the swap Monte Carlo algorithm. Our study highlights the potential of generative machine learning models in scientific computing for complex systems, but also points to some of its current limitations and the need for further improvement.
\end{abstract}

\maketitle

\section{Introduction}

One of the most important methodological revolution in science in the last century is scientific computing~\cite{battimelli2020computer}. Numerical simulations represent a way, complementary to experiments, to study physical systems, thus providing a unique lens on the microscopic mechanisms underpinning macroscopic physical phenomena~\cite{battimelli2020computer,frenkel2001understanding,allen2017computer,newman1999monte}. 

A major application of numerical simulations, from the very beginning, has been sampling physical configurations at thermal equilibrium~\cite{metropolis1953equation}. This was initially done by using either Monte Carlo Markov chains~\cite{metropolis1953equation,hastings1970monte} or molecular dynamics~\cite{alder1957phase,rahman1971molecular}. Both methods can be viewed as ways to implement some physical dynamics to ergodically explore the configuration space, just as the physical system does. The basic challenge is to run those dynamics long enough to be able to generate a large set of uncorrelated configurations to perform accurate ensemble averages of physical observables~\cite{frenkel2001understanding}.

When the system is characterized by large relaxation times, for instance near phase transitions or in disordered media, sampling can become so challenging that conventional methods may fail~\cite{frenkel2001understanding,newman1999monte}. In such cases, the only solution, so far, consists in devising alternative dynamics that ensure equilibrium sampling while being characterised by substantially smaller decorrelation times. Such strategies are described in many standard textbooks on computer simulations and statistical physics~\cite{newman1999monte,allen2017computer,krauth2006statistical,landau2021guide}. In the context of off-lattice molecular simulations, we can mention non-local~\cite{swap:GAZZILLO1989,kranendonk1991computer,grigera2001fast}, lifting~\cite{vucelja2016lifting}, or collective~\cite{bernard2009event,krauth2021event,ghimenti2024irreversible} Monte Carlo algorithms, parallel tempering~\cite{marinari1992simulated,hukushima1996exchange,earl2005parallel,swendsen1986replica}, population annealing~\cite{hukushima2003population,machta2010population,amey2018analysis}, or irreversible Langevin dynamics~\cite{ghimenti2023sampling}. 

However, there exist physical systems in which equilibrium sampling remains a difficult open problem even with enhanced sampling methods. Among them, glassy systems~\cite{berthier2011theoretical} stand as one of the most difficult to simulate in condensed and soft-matter physics. Molecular, colloidal and spin glasses are in fact known to display an extremely slow physical dynamics, which creates a major challenge to standard simulation algorithms~\cite{berthier2023modern,barrat2023computer}. Glassy systems can in fact serve as a severe test of any newly proposed method, and can be seen as a paradigm for complex systems.  

The recent discovery of generative models in artificial intelligence (AI) able to generate large structured data such as images, sound, 3D-video, and text has the potential to induce a second revolution in scientific computing~\cite{goodfellow2014generative,sohl2015deep,ho2020denoising,kingma2013auto,rezende2015variational}. These AI models are not only able to accurately produce complex data, but are also very fast. Speed is a central requirement in the realm of scientific computing. Several applications appeared already.  In 2019, Noé \emph{et al.} proposed the usage of normalizing flows (NF)~\cite{noe2019boltzmann}, and independently Wu \emph{et al.} variational autoregressive models~\cite{wu2019solving}, for Boltzmann sampling in statistical and condensed matter physics. These works have found numerous applications for sampling~\cite{invernizzi2022skipping,gabrie2022adaptive,falkner2023rare,coretti2022learning,vanleeuwen2023boltzmann} and free energy calculations~\cite{xinqiang2020free,wirnsberger2023free}. However, despite these interesting premises, a clear view on when, where and how these methods work, and in particular their limitations and efficiency against known algorithms, is currently lacking. For standard phase transitions, promising results have been obtained in \cite{marchand2022wavelet,singha2023conditional}. However, for hard computational problems such as complex and glassy systems, it is unclear whether they can circumvent the problem of large relaxation times. Positive results have been reported in \cite{mcnaughton2020spin,scriva2023spinglass} for spin-glasses. On the other hand, theoretical and numerical analysis of mean-field models for structural glasses~\cite{ciarella2023generative,ghio2023sampling}, related to other hard problems in computer science~\cite{mezard2009information}, has shown that several generative models do not, and sometimes can not, have good performances. Worse, they sometimes perform less efficiently than conventional algorithms, such as local Monte Carlo~\cite{ciarella2023generative}. These results provide a rather pessimistic view of the potentiality of machine learning (ML) techniques in the field of glassy systems. A last difficulty is that the performance of generative models is generically expected to scale very badly with the size of the system, so that applications to study phase transitions and collective effects in many-body systems appear out of sight.        

Given the rapid progress made in ML studies~\cite{ronhovde2012detection,schoenholz2016structural,bapst2020unveiling,paret2020assessing,jung2023roadmap}, we feel that there is room for hope and progress. At the moment, there is a clear need of further studies to develop and test generative models in hard computational problems, and benchmark their performances against the ones of existing algorithms. The aim of this work is to perform such analysis for atomistic models of glass-forming liquids~\cite{berthier2023modern}. Contrary to \cite{ciarella2023generative,ghio2023sampling}, we study an off-lattice, finite-dimensional glassy model which displays an extremely slow dynamics, associated to a super-Arrhenius evolution of relaxation and sampling times. It is challenging to perform numerical simulations in realistic experimental conditions for this model, as the physical relaxation time increases by more than 14 orders of magnitude towards the experimental glass transition temperature~\cite{PhysRevX.12.041028}.

We focus on a specific two-dimensional glass-forming model that shows conventional signatures of glassy dynamics~\cite{jung2022predicting,jung2024dynamic}, and represents therefore a relevant and challenging test bench for enhanced sampling methods. At low temperatures, molecular dynamics becomes rapidly unable to perform an equilibrium sampling of the configuration space, even for modest system sizes. Given the challenges mentioned above, we intentionally study a relatively small system size to separate the capabilities of NF to tackle complex landscapes from its potentially problematic scaling with system size.  

We optimise a ML technique based on normalizing flows~\cite{rezende2015variational,noe2019boltzmann}, which we carefully benchmark against several advanced techniques introduced in the physics literature, such as parallel tempering~\cite{hukushima1996exchange}, population annealing~\cite{machta2010population}, and swap Monte Carlo~\cite{swap:ninarello2017}. By studying and comparing their abilities to produce an ensemble of equilibrated low-temperature configurations, we provide the first quantitative analysis of the performance of ML methods to sample realistic supercooled liquids at low temperatures. Surprisingly, our results demonstrate the great potential of such method which turns out to be much more efficient than conventional molecular dynamics and achieves performances comparable to parallel tempering and population annealing. Finally, we assess current limitations of these new methods and provide guidelines for further studies, in particular to improve the parametrization of the normalizing flow and to extend this technique to larger system sizes.

The paper is organised as follows.
In Sec.~\ref{sec:model} we define the numerical glass-forming model and explain how to assess the performance of sampling algorithms. 
In Sec.~\ref{sec:benchmark} we benchmark various known algorithms: molecular dynamics, swap Monte Carlo, parallel tempering and population annealing. 
In Sec.~\ref{sec:NF} we introduce, optimize, and study the performance of a NF model. 
Finally, in Sec.~\ref{sec:discussion} we collect our results, and discuss the implications for future research. 

\section{Setting the Stage: Model and sampling task}

\label{sec:model}

\subsection{A two-dimensional ternary Lennard-Jones mixture}

We study a two-dimensional ($d=2$) model introduced and developed in Ref.~\cite{jung2022predicting}. This is a variation of the binary Lennard-Jones mixture introduced long ago by Kob and Andersen~\cite{kob1995KA} in which a third component is introduced to both improve the glass-forming ability (i.e. to prevent easy crystallization) and enable a more efficient use of the swap Monte Carlo algorithm, a strategy proposed in \cite{swap:ninarello2017,Berthier2020}. We refer to \cite{jung2022predicting,jung2024dynamic} for all details regarding the model parameters and simulation details. 

\begin{figure}[b]
    \centering
    \includegraphics[scale=0.25]{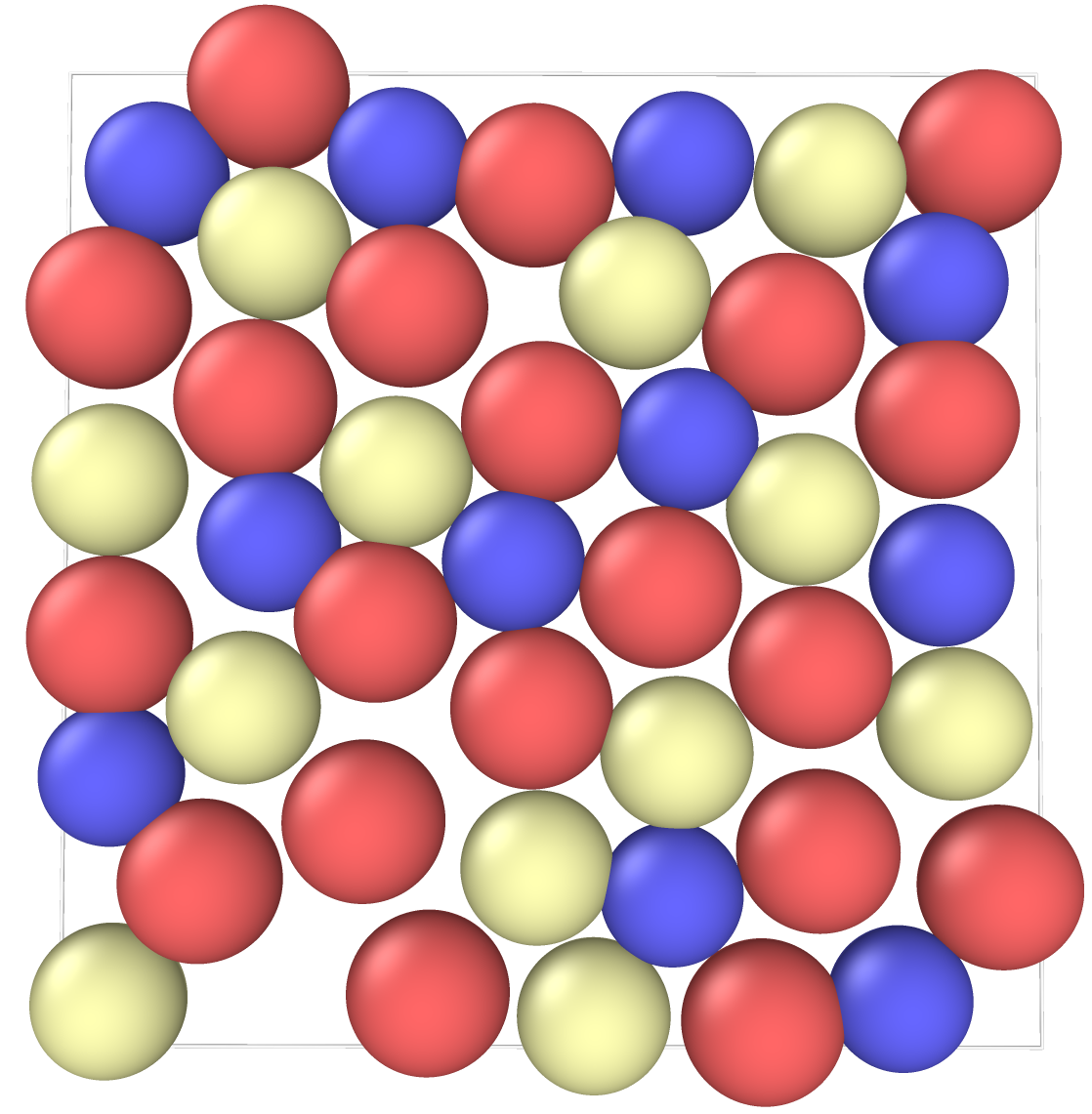}
    \caption{\gj{Snapshot of a typical amorphous glassy configuration of the two-dimensional model at temperature $T=0.205.$ Colors indicate different types of particles. The goal of this study is to produce a large number of independent configurations drawn from the Boltzmann distribution in Eq.~(\ref{eq:boltzmann}).}}
    \label{fig:snapshot}
\end{figure}

We investigate systems with $N=43$ particles using periodic boundary conditions with box length $L=6.0$, in reduced units \gj{(see Fig.~\ref{fig:snapshot} for a snapshot)}. The unit of length is $\sigma$, which corresponds to the diameter of the large particles. When using molecular dynamics, the unit of time is the Lennard-Jones timescale $\tau = \sqrt{m \sigma^2/\epsilon}$ where $m$ is the particle mass, and $\epsilon$ the interaction strength between large particles. For different algorithms, we express times in units of $\tau$, in order to carefully reflect the actual computational cost of each method.

The relatively small system size is actually comparable to previous studies using normalizing flows for sampling in complex systems~\cite{noe2019boltzmann,invernizzi2022skipping,coretti2022learning}. \gj{The total dimensionality of the problem is $D= Nd = 86 $}.  The main goal of this work is to benchmark the efficiency of normalizing flows in \gj{sampling such small glassy systems according to the Boltzmann distribution,}
\begin{equation}\label{eq:boltzmann}
    \rho_*(x) = Z_*^{-1} \exp( - \beta_* U(x)).
\end{equation}
Here, $Z_*$ is a proportionality constant, $\beta_*$ is the target inverse temperature  and \gj{$U(x)=E_\text{pot}(x)$} is the total (potential) energy. For each configuration, we measure the total potential energy, $E_{\rm pot} = \sum_{i \neq j} V(r_{ij})$, where $V(r)$ denotes the short-range repulsive pair interaction potential, \gj{as defined in the Supp. Mat. I of Ref.~\cite{jung2022predicting}}, and $r_{ij}$ the relative distance between particles $i$ and $j$.

Results for a larger system size, $N=172$, are presented in Appendix~\ref{app:large}. Scaling to larger systems introduces additional challenges that have to be considered separately and are left for future work. While $N=43$ seems a small number of particles, we emphasize that this is large enough to produce glassy dynamics and local structure for this dense fluid that are nearly equivalent to those of much larger systems~\cite{heuer2008exploring} (see also Fig.~\ref{fig:snapshot}). This implies in particular that equilibrium sampling even for this modest system size is already a difficult computational challenge.  

\subsection{The specific heat as a sampling task}

\label{sec:sampling_task}

In order to assess and compare the properties of various algorithms, we first need to define a specific sampling task to be able to test how well and how fast that task is achieved by the various algorithms.

In supercooled liquids approaching their glass transition, changes in many structural quantities are typically very modest, and deciding whether or not a given configuration is equilibrated is not straightforward. The standard solution is to measure time correlation functions, as glassy dynamics is extremely sensitive to small temperature changes, so that lack of equilibration, insufficient sampling, or small drifts are more easily detected using dynamic quantities. This approach is however not available to parallel tempering, population annealing and normalizing flows which output a set of low-temperature configurations that are not connected by any obvious dynamics.

To solve this problem we analyse the statistics of energy fluctuations measured in an ensemble of configurations.  We define the average potential energy over this ensemble, $\langle E_{\rm pot} \rangle$, and the variance of its fluctuations, which is directly connected to the specific heat as \cite{allen2017computer}
\begin{equation}
    c_V = \frac{C_V}{N} = \frac{\langle E_{\rm pot}^2 \rangle - \langle E_{\rm pot} \rangle^2}{Nk_B T^2}  .
\end{equation}
\gj{Here, $T$ denotes the temperature of the system and $k_B$ the Boltzmann constant ($k_B=1$ in our units).}
\gj{A correct estimate of the specific heat at a given temperature thus requires the production of several independent equilibrium configurations in order to correctly assess the fluctuations around the mean $\langle E_{\rm pot} \rangle$.} We conclude that the determination of the specific heat represents a well-defined task that is able to probe the capability of a given algorithm to (i) reach thermal equilibrium, (ii) sample a large number of independent configurations $x$ representative of the Boltzmann distribution.
Another advantage is that $c_V$ does not require knowledge of a dynamics between configurations, and is thus broadly applicable to any sampling technique. Due to the generality of this sampling task we therefore believe that it is similarly suited to benchmark enhanced sampling techniques for various different complex systems.

In practice, we additionally define a convergence timescale which quantifies the computational time it takes for a given algorithm to correctly approach the equilibrium value of the energy at a given temperature. This timescale will thus allow us to rank the different algorithms by their efficiency to accomplish the requested sampling task. 

We also studied alternative, previously-proposed determination of equilibration, such as different definitions for the specific heat related by a fluctuation-dissipation relation, the radial distribution function, histograms of potential energies, and density of states. See Appendix \ref{app:equilibrium} for more details on these other approaches. We found that none of these measures can reliably be used, as they often overestimate the degree of equilibration and are blind to small deviations from equilibrium. Therefore, we focus on $c_V$ as our main observable.

\section{Benchmarking known sampling algorithms}

\label{sec:benchmark}

In this section we analyse the sampling performances of four distinct algorithms: swap Monte Carlo (SMC), molecular dynamics (MD), parallel tempering (PT) and population annealing (PA).  

\subsection{Swap Monte Carlo (SMC)}

A major problem for benchmarking enhanced sampling techniques is usually the absence of a reference solution and therefore of a clear performance measure. Here, this issue is easily settled by using swap Monte Carlo (SMC)~\cite{grigera2001fast,berthier2016equilibrium,swap:ninarello2017}. The algorithm adds non-local Monte Carlo moves on top of conventional molecular dynamics simulations~\cite{swap:Berthier2019}. The Monte Carlo moves are swap moves, in which a pair of particles with different types are randomly selected and their radii are swapped. This swap move is accepted according to a Metropolis scheme. This algorithm is extremely efficient and can be used to equilibrate supercooled liquids at extremely low temperatures, including below the experimental glass transition temperature~\cite{swap:ninarello2017,Berthier2020,jung2022predicting}. \gj{A detailed introduction and discussion of this algorithm can be found in Ref.~\cite{swap:Berthier2019}.}

\begin{figure}
    \centering
    \includegraphics[width=\linewidth]{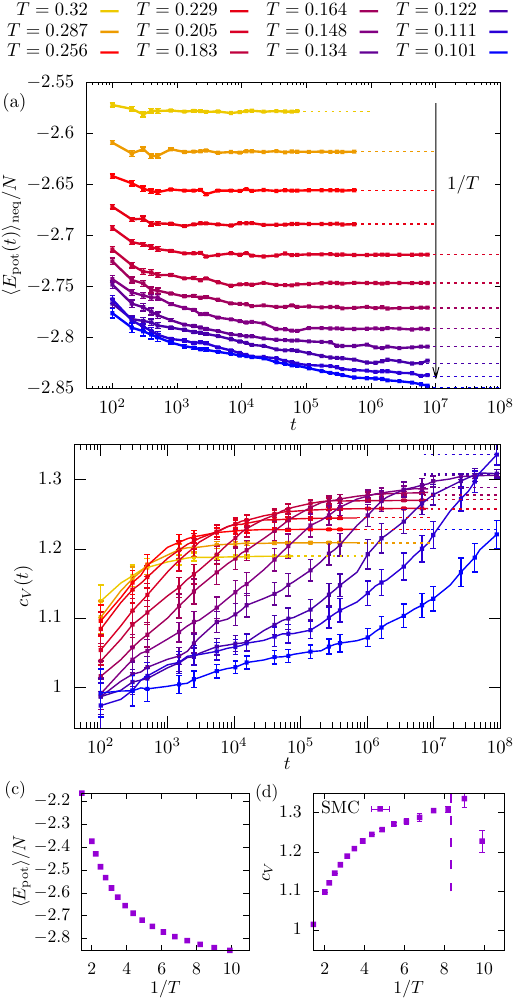}
    \caption{{\bf Sampling with swap Monte Carlo (SMC).} (a) Equilibration of the potential energy $\langle E_\text{pot}(t)\rangle_\text{neq}$ ($T_\text{init}=0.5$ at $t=0$). (b) Sampling of the specific heat $c_V(t)$, which no longer reach a plateau for temperatures $T < 0.12 < T_g.$ Horizontal dashed lines show the long-time averages.
    (c) Long-time average of the potential energy $\langle E_\text{pot} \rangle$ and (d) the specific heat $c_V$. The vertical line marks $T_\text{SMC}=0.12.$ which is the temperature below which SMC sampling fails. } 
        \label{fig:SMC}
\end{figure}

Anticipating that SMC is the most efficient sampling method, we therefore perform SMC simulations to provide a benchmark for the following analysis of other sampling methods. In detail, we perform 105 swap attempts every 50 MD steps. This is the highest swap frequency that we could use without inducing small, but noticeable energy shifts, which were then affecting the quality of the benchmarking performed below. Since swap moves are not frequent, it is pertinent to use the Lennard-Jones MD time unit $\tau$ as the time unit also for the SMC method. 

We first equilibrate the system for roughly $10^5 - 10^7 \,\tau$ starting from temperature $T_\text{init}=0.5.$ During equilibration, we monitor the non-equilibrium potential energy, $\langle E_\text{pot}(t)\rangle_\text{neq} $. Here, the average, $\langle\cdots\rangle_\text{neq}$, is taken over $N_s=64$ independent simulations, all starting from different equilibrium configurations sampled at $T_\text{init}$ at $t=0.$ Afterwards, we perform SMC sampling for another $t_\text{samp}=10^7 - 10^8 \,\tau$ (depending on the temperature) to extract the ensemble average as introduced in Sec.~\ref{sec:sampling_task}. During sampling, we also extract the time-dependent average $ \langle X \rangle_t = \sum_{t_s < t} X(t_s) $, from which we obtain the time-dependent specific heat $c_V(t) = (\langle E_{\rm pot}^2 \rangle_t - \langle E_{\rm pot} \rangle_t^2 ) / (NT^2).$ In the limit $t \rightarrow t_\text{samp}$ we then recover the long-time average $ \langle X \rangle_t \rightarrow \langle X \rangle$. We have ensured that the measured timescale reflects the actual computational cost for these SMC simulations to enable quantitative comparison of equilibration and sampling timescales. 

Results for the time dependence of different observables during equilibration and then during sampling are shown in Fig.~\ref{fig:SMC}. The potential energy $\langle E_\text{pot}(t)\rangle_\text{neq} $ decays strongly during equilibration until it reaches a plateau. Only for temperatures significantly below the estimated glass transition temperature ($T_g \approx 0.15$) we observe that the potential energy continues to decay even beyond $t > 10^7$, suggesting that SMC falls out of equilibrium at these temperatures. 

We also investigate the time dependence of the specific heat measured after the long equilibration run. Here and in the following, error bars are calculated from the variance over several independent runs. Starting from a small value at short time (when a single configuration has been probed), $c_V(t)$ rapidly accumulates on short time scales contributions from vibrations within one state (leading to $c_V \approx 1$, since we work in $d=2$ space dimensions). At much later times, the system visits a manifold of different states to eventually correctly sample the Boltzmann distribution. Different from the equilibration discussed above, reaching a plateau in $c_V(t)$ requires much longer times for $c_V$ than for $\langle E_{\rm pot} \rangle$: it takes longer to explore enough configurations to estimate $c_V$ than simply reaching an energy value close to the equilibrium one.

Explicitly, we can deduce from Fig.~\ref{fig:SMC}(b) that the system does not reach a plateau anymore for temperatures $T<T_\text{SMC}=0.12$ during sampling of $c_V(t)$. We therefore identify $T_\text{SMC}$ as the temperature below which SMC is no longer able to perform the assigned sampling task. Notice that near $T_{\rm SMC}$ the energy can reach a plateau, indicating that the system is very close to equilibrium, but the simulations are nevertheless not sufficiently long to sample a large enough number of independent configurations to provide the correct estimate of the specific heat. 

The corresponding long-time averages of $\langle E_\text{pot} \rangle$ and $c_V$ are shown in Fig.~\ref{fig:SMC}. We observe that the potential energy decays monotonically for decreasing temperature. Furthermore it can be seen that above $T > T_\text{SMC}$ the results for $c_V$ are monotonically increasing with decreasing temperature. Once the system falls out of equilibrium near $T_{\rm SMC}\approx 0.12$, there is a strong decrease of $c_V$, as found previously in many experiments and simulations in cases where the system falls out of equilibrium~\cite{flenner2006hybrid}.

In the following, we will consider the SMC results down to $T_{\rm SMC}$ as reflecting the correct equilibrium behavior, in order to test and benchmark the performance of the alternative techniques.

\subsection{Molecular dynamics (MD)}

\label{sec:MD}

\begin{figure}
    \centering
    \includegraphics[width=\linewidth]{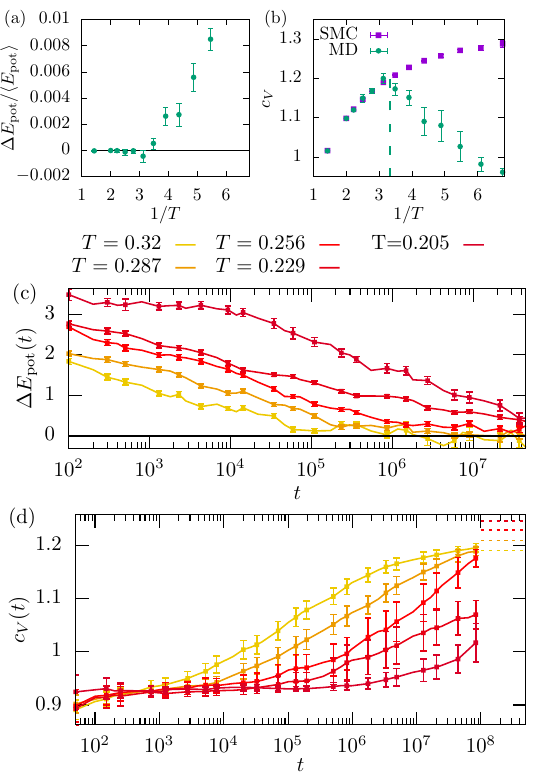}
    \caption{{\bf Benchmarking MD dynamics.} (a) Difference to the SMC potential energy, $\Delta E_\text{pot} = \langle E_\text{pot}^\text{MD} \rangle - \langle E_\text{pot}^\text{SMC} \rangle,$ where $\langle E_\text{pot}^\text{MD} \rangle$ is the long-time average of the MD dynamics.  (b) Long-time average of the specific heat $c_V.$ The vertical lines mark $T_\text{MD}=0.3,$ below which MD sampling fails. (c) Equilibration of the potential energy characterized by the quantity $\Delta E_\text{pot}(t) = \langle E_\text{pot}^\text{MD}(t)\rangle_\text{neq} - \langle E_\text{pot}^\text{SMC} \rangle.$ (d) Sampling of the specific heat $c_V(t).$ $c_V(t)$ does not reach a plateau anymore for temperatures $T <  T_\text{MD}.$ Dashed horizontal lines mark the long-time SMC results. Color code in (c, d) as in Fig.~\ref{fig:SMC}. }
    \label{fig:MD}
\end{figure}

Molecular dynamics (MD) simulations consist in solving Newton's equations of motion with an added thermostat to control the temperature~\cite{allen2017computer}. Consequently, the dynamic relaxation proceeds through realistic dynamics. Our simulations use a Nose-Hoover thermostat with relaxation time $\tau_\text{NH}=1.0$ and time step of $\Delta t = 0.005.$ Identical to SMC we create high temperature configurations at $T_{\rm init}=0.5$ and then quench the temperature to the desired value $T$ to monitor the relaxation of the potential energy towards equilibrium. After thermalization for $t > 10^7 \,\tau$, we investigate $\langle E_\text{pot} \rangle_t$ and its fluctuations to measure $c_V(t)$. In order to perform a fair comparison, we use the same range of timescales and report the same averaged quantities for MD, SMC and all other sampling methods.

Results for MD dynamics are shown in Fig.~\ref{fig:MD}. Since SMC provides equilibrium measurements down to low temperatures, we can study the difference $\Delta E_\text{pot} = \langle E_\text{pot}^\text{MD} \rangle - \langle E_\text{pot}^\text{SMC} \rangle$ to better quantify differences to the established SMC results which are equilibrated down to $T_\text{SMC}$. For MD, the potential energy starts to systematically deviate from the expected SMC result already for temperatures $T < 0.3$, see Fig.~\ref{fig:MD}(a). This is confirmed in Fig.~\ref{fig:MD}(b) which shows that the specific heat measured by MD simulations shows a peak near $T_{\rm MD} =0.3$, indicating lack of sampling for lower temperatures. 

While equilibrium dynamics can easily be measured for MD using for instance time correlation functions, we show instead how the energy decay after a quench from $T_{\rm init}=0.5$ and the time dependence of the specific heat measured during equilibrium sampling in Figs.~\ref{fig:MD}(c,d). Compared to SMC, we show a narrower regime of temperatures down to $T=0.205$. In Figs.~\ref{fig:MD}(c) we observe an increasing time scale to reach the correct value of the energy which becomes impossible for $T<T_{\rm MD}$ over the simulated time window. The lack of sampling becomes more severe when considering the specific heat which can only reach its plateau value for $T=0.32$ but not below. Our data confirm that MD is much less efficient than SMC, as expected. More importantly perhaps, since MD follows the physical dynamics of the system, the data in Fig.~\ref{fig:MD} in fact serve as a benchmark in order to assess how much gain over the physical dynamics any enhanced algorithm can achieve~\cite{ghimenti2024irreversible}. 

\subsection{Monte Carlo in temperature space: Parallel tempering (PT)}

\label{sec:PT} 

Parallel tempering \cite{earl2005parallel}, also known as replica exchange \cite{swendsen1986replica,hukushima1996exchange}, is a popular enhanced sampling technique applied in a wide range of fields, including spin glasses \cite{swendsen1986replica,hukushima1996exchange}, protein folding \cite{sugita1999replica, bussi2006free}, polymer melts \cite{PhysRevE.63.016701}, and solid state physics \cite{falcioni1999biased}. In the field of glass-forming liquids, it has been used to create equilibrium structures of deeply supercooled liquids \cite{PhysRevE.61.5473,de2002equilibration}, characterize point-to-set length scales \cite{yaida2016point,berthier2016efficient}, and analyze the physics of randomly pinned systems~\cite{kob2013probing}. 

The key idea is to perform several MD simulations in parallel, each using the same MD parameters as explained in Sec.~\ref{sec:MD} but running with a set of $n$ different temperatures $T_0, \ldots T_{n-1}$. Each of these simulations is called a replica. In addition to MD steps, every $N_\text{ex}$ MD steps we attempt to exchange the configurations between two replicas with adjacent temperatures. The exchange of the configuration in replica $j$ with the one in the neighbor $j \pm 1$ is accepted according to a Metropolis scheme,
\begin{equation}
P_\text{acc}(j \leftrightarrow j \pm 1) = \exp( - (\beta_j - \beta_{j \pm 1}) \Delta U ),
\end{equation}
with energy difference \gj{$\Delta U = E_\text{pot}(x_j) - E_\text{pot}(x_{j\pm1})$} between the two configurations and inverse temperatures $\beta_j= (k_B T_j)^{-1}$. \gj{An extended derivation of this equation and efficient implementation can be found in Chapter 14.1 of Ref.~\cite{frenkel2001understanding}.} Using this algorithm configurations evolve both by the physical MD dynamics but also by performing a random walk in temperature space. Low-temperature configurations can therefore follow an ``easy'' relaxation path by being exchanged with replicas from higher temperatures, then evolving faster at these high temperatures and subsequently being exchanged back to the low temperature. This may avoid configurations being stuck for extremely long times in deep minima at low temperatures. Physically, if the basins relevant at low temperature are also sampled at high temperature, PT can become a very efficient method~\cite{berthier2023modern}.

The most important factor to optimize PT simulations is the choice of the replica temperatures. On the one hand, large temperature differences will significantly reduce the acceptance rate for exchange events and therefore slow down the exploration of the temperature space. On the other hand, a large number of replicas implies a larger computational effort. After trial and error, we finally use $n=8$ replicas with temperatures $T=$ $0.4$, 0.359, 0.32, 0.287,0.256,0.229, 0.205, $0.183$. \gj{To arrive at this choice, we have started with a small number of replicas and systematically increased their number until we found the optimal result in the given computational time.} This choice is rationalized by a significant overlap between energy distributions at neighboring temperatures and therefore a large acceptance rate $\langle P_\text{acc} \rangle > 0.25$ for all replicas. We also checked that smaller temperature steps do not improve the results. We optimized the maximal and minimal temperatures in the above range to finally settle on this list of $n=8$ replicas. We also choose $N_\text{ex}=5000$ which is a reasonable choice between too frequent or too infrequent temperature swaps, but the results are not very sensitive to this specific choice. 

\begin{figure}
    \centering
    \includegraphics[width=\linewidth]{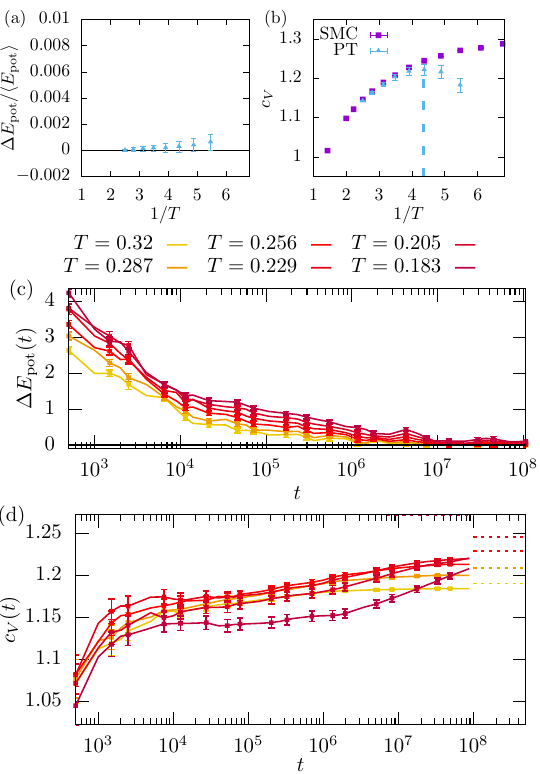}
    \caption{{\bf Benchmarking parallel tempering (PT) simulations.} 
    The description is the same as for Fig.~\ref{fig:MD}. The vertical dashed line in (b) represents $T_{\rm PT}=0.23$ below which PT sampling fails.}
    \label{fig:PT}
\end{figure}

The results for the benchmarking of PT are shown in Fig.~\ref{fig:PT} using the same organization as for MD. Comparing to Fig.~\ref{fig:MD} shows that PT is superior to MD. Within errorbars, PT predicts the correct potential energy in a temperature regime in which MD is already substantially out of equilibrium. Using the specific heat as a sharper test for sampling in Fig.~\ref{fig:PT}(b), we conclude that PT succeeds in the sampling task down to $T_{\rm PT}=0.23$, below which $c_V$ decreases as a result of insufficient sampling. This temperature is considerably lower than $T_{\rm MD}=0.3$, but much higher than $T_{\rm SMC}=0.12$.

To understand better the efficiency and limits of the PT sampling we turn to the energy decay after a quench from an initial condition where all $n$ replicas are initialized at a high temperature $T_{\rm init}=0.5$, see Fig.~\ref{fig:PT}(c). Interestingly, the relaxation time of $\Delta E_\text{pot}(t)$ is only weakly dependent on the temperature. This behavior is qualitatively different from the MD results. Therefore, while PT accelerates dynamics at low temperatures, it also slows down the dynamics at higher temperatures compared to MD. This is the direct result of the nature of PT exchange events: since replicas travel across the entire temperature spectrum, there is a nearly unique emerging timescale controlling the approach to equilibrium of the entire simulation composed of $n$ replicas. In other words, different temperatures are no longer independent when using PT, and the relaxation is in effect slaved to the slowest replica. This conclusion also explains why we did not include replicas with even lower temperatures into the PT dynamics as it negatively impacts the performance of the PT simulations.  

The time dependence of the specific heat, $c_V(t)$, is more interesting, see Fig.~\ref{fig:PT}(d). Here, the equilibrium plateau in $c_V$ is reached faster for higher temperatures, and the corresponding timescale becomes longer than our simulation time for $T< T_{\rm PT}$, explaining the spurious peak in the measured specific heat in Fig.~\ref{fig:PT}(b). The time dependent relaxation can thus be used to quantify the speedup offered by PT simulations over MD, and this will be discussed in Sec.~\ref{sec:quantitative}.

\subsection{Population annealing (PA) and reweighting (RW)}

\label{sec:PA}

Population annealing (PA) is deeply rooted in the reweighting (RW) technique known from statistical mechanics, which we recap first.

Given a set of $R$ configurations, $\{ x_i\}$, with $i=1,...,R$ taken from the Boltzmann $NVT$ ensemble at temperature $T_1$, it is possible to reweight these configurations to perform an equilibrium average at a different temperature $T_2$~\cite{ferrenberg1988new}. To this end, one assigns a new Boltzmann weight $W_i= \exp (- (\beta_1 - \beta_{2}) U(x_i) )$ to each configuration $i$. The ensemble average of an observable $A(x)$ at $T_{2}$ is given by 
\begin{equation}
    \langle A \rangle_{T_{2}} = \frac{\sum_i W_i A(x_i)}{ \sum_i W_i }.
\label{eq:RW}
\end{equation}
Reweighting is used extensively for free energy calculations in molecular simulations~\cite{shen2008statistical,miao2014improved}. The method, however, only works efficiently for small enough temperature steps so that the weights $W_i$ remain meaningful.  
 
In PA, a large set of configurations is used to perform small temperature steps to gradually anneal the temperature to the target low-temperature, while reweighting their Boltzmann weights at each step~\cite{hukushima2003population,tokdar2010importance,machta2010population,PhysRevE.92.063307,PhysRevE.97.033301,gessert2023resampling}. In practice, the PA algorithm works as follows. We first create $R$ configurations at an initial, high temperature $\beta_1 = (k_B T_1 )^{-1}$ and evaluate the Boltzmann weight of each configuration $x_i$ as $W_i= \exp (- (\beta_1 - \beta_{2}) U(x_i) )$. Here, $T_{2} $ should be slightly smaller than $T_1$. We then create on average $\tau_i$ copies of each configuration $i$, where $\tau_i$ is given by
\begin{equation}\label{eq:copies}
	\tau_i = R \frac{W_i}{\sum_{k=1}^R W_k}. 
 \end{equation}
Recently, different schemes were compared to numerically implement Eq.~(\ref{eq:copies})~\cite{gessert2023resampling}. We apply the ``systematic resampling'' scheme, which was the most efficient for a constant population size $R$. Following resampling, we finally perform $M$ MD steps on each copy at temperature $T_2$ to help thermalize the configurations at the new temperature $T_2$. This ends the annealing from $T_1$ to $T_2$. This annealing step $T_1 \to T_2$ is then repeated several times until the final target temperature $T$ is reached. Each annealing step consists in (i) resampling the population (ii) a small number of $M$ MD steps for each configuration. \gj{More extended derivations of the technique and algorithms can be found in Ref.~\cite{machta2010population}.}

For the choice of annealing temperatures in PA, we use the same series $T_1, \ldots, T_{n-1}$ used for PT in Sec.~\ref{sec:PT}. This is reasonable since the PT temperatures were optimized to provide good overlaps between the probability distributions of potential energy, which also controls the quality of the reweighting in Eq.~(\ref{eq:RW}). Contrary to PT, the annealing procedure in PA is unidirectional as the population flows from $T_1$ to $T_{n-1}$ but no information is carried backwards. As a result, including lower temperatures is harmless (at worse, PA sampling fails) and so we include two lower temperatures $T_{n}=0.164$ and $T_{n+1}=0.148$. We perform high temperature MD simulations at $T_1 = 0.359$ and save configurations every $10^4 \,\tau,$ which corresponds roughly to the MD structural relaxation time at this $T_1$. This choice ensures that within a similar computational effort invested into PT we can create an initial set of $R= 2\times 10^5$ statistically independent configurations. In addition, this comparison enables us to assign a computational time $t = R \times 10^4 \tau$ to the PA task, as the annealing steps themselves can be efficiently performed. None of the above choices critically affects the result when reasonably changed. The most critical parameter is the number $M$ of MD relaxation steps. A too small number $M < 10^3$ leads to tiny but systematic differences in the observed $\langle E_\text{pot} \rangle$ and $c_V.$ We therefore choose $M=5 \times 10^3$. Since the creation of the initial set of $R$ configurations is the computational bottleneck, such a large $M$ value does not significantly increase the computational effort.

\begin{figure}
    \centering
    \includegraphics[width=\linewidth]{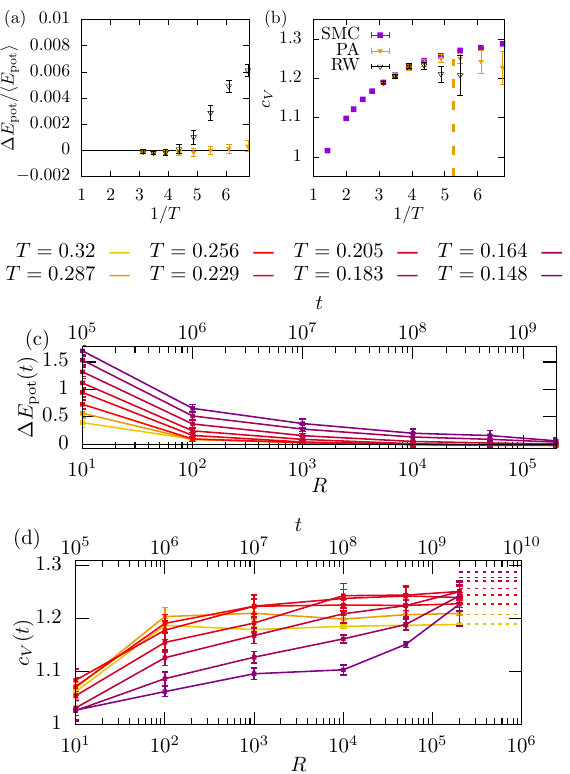}
    \caption{{\bf Benchmarking population annealing (PA) and reweighting (RW).} The description is the same as for Fig.~\ref{fig:MD}. The time dependence shown in panels (c) and (d) is obtained by using different numbers of initial configurations $R.$ The vertical dashed line in (b) represents $T_{\rm PA}=0.19$ below which PA sampling fails.}
    \label{fig:PA}
\end{figure}

It is instructive to compare the gradual population annealing from $T_1$ to a given target temperature $T$ with a direct reweighting performed in a single step $T_1 \to T$ directly using Eq.~(\ref{eq:RW}) applied to the entire initial population of configurations created at $T_1$, see Fig.~\ref{fig:PA}. We observe that RW is already much more efficient than MD dynamics with correct energy and specific heat obtained down to $T_{\rm RW}=0.25$. The effect of the gradual annealing and resampling performed within PA improve the RW results dramatically, and nearly-correct energy values are predicted down to the lowest temperature. A more careful inspection of the $c_V$ data shows however that PA sampling fails below $T_{\rm PA}=0.19$.   

Different from MD and PT, there are no separate equilibration and sampling procedures within PA. Nevertheless, it is possible to provide an equivalent time dependence to both $\Delta E_\text{pot}(t(R))$ and $c_V(t(R))$ by following the evolution of the PA performance as a function of the population size $R$ produced at the initial temperature $T_1$, because by definition correct sampling is obtained in the limit $R\to\infty$, just as correct sampling is performed in the large time limit for any of the other algorithms. \gj{Convergence in the large time or population limit is obvious, since system size and thus energy barriers are finite. However, our problem is to reach convergence in a tractable computational timescale.} We can then convert the population size $R$ into a computational timescale using the dictionary $t = 10^4 R$, which corresponds to the effective computational time invested into creating the set of $R$ configurations.

The results are shown in Figs.~\ref{fig:PA}(c,d) which illustrate the convergence of the energy and its fluctuations to the correct values as $R$ is increased. Differently from PT, we see that the equilibration of the potential energy slows down with temperature, see  Fig.~\ref{fig:PA}(c). We can still observe the relaxation of $c_V(t)$ towards its equilibrium plateau value. In fact the $R$ dependence of $c_V$, and its eventual convergence to a plateau at large $R$, serves as a stringent test of the quality of sampling with PA. In particular, we confirm that PA sampling fails below $T_\text{PA}=0.19$. This result shows that PA performs slightly better than the two previous sampling methods: $T_\text{PA} < T_\text{PT} < T_\text{MD}$, a result that could not be anticipated based on previous efforts. An additional advantage of PA is that the task of sampling an initial set of $R$ independent samples at the high temperature $T_1$ can be easily parallelized, by running several independent simulations in parallel.

\section{Sampling by Normalizing Flows}

\label{sec:NF}

In Sec.~\ref{sec:benchmark} we established benchmarks for enhanced sampling techniques known from physics. This sets the stage for a thorough analysis of the performance of the ML method of normalizing flows (NF)~\cite{rezende2015variational,papamakarios2021normalizing}. Since NF are relatively new methods, we first provide a general introduction before describing our implementation and the main results. 

\subsection{Continuous normalizing flows (NF)}

The general idea of normalizing flows is to learn an invertible mapping between two probability distributions: a prior distribution $\rho_P(x)$, from which we can sample easily (Gaussian random numbers, high temperature liquids), and a target distribution, $\rho_*(x)$, in our case the Boltzmann distribution~\cite{noe2019boltzmann}. The mapping is in general only approximate, so one needs to reweight the configurations obtained by the NF. A more accurate mapping leads to lower rejection. Normalizing flows found applications in computer vision~\cite{dinh2016density}, sampling via Markov Chain Monte Carlo~\cite{song2017nice,gabrie2022adaptive,klein2024timewarp}, lattice field theories~\cite{albergo2021introduction} and condensed matter physics~\cite{noe2019boltzmann,vanleeuwen2023boltzmann}. 

Boltzmann generators are the first application of normalizing flows for sampling of complex systems in condensed matter \cite{noe2019boltzmann}. Compared to applications such as image generation, a specific property of Boltzmann generators is that the target distribution is known and corresponds to the Boltzmann distribution (see Eq.~\ref{eq:boltzmann}). The challenge for Boltzmann generators in statistical physics is to efficiently sample from this distribution using the learned NFs. For a general introduction to Boltzmann generators see Refs.~\cite{noe2019boltzmann,coretti2022learning}. 

Here, we use equivariant, continuous normalizing flows~\cite{kohler2020equivariant}. This specific NF has the advantage of being equivariant to translations, rotations \gj{and permutations} and thus mirrors the fundamental symmetries of the \gj{underlying physical system. In Ref.~\cite{kohler2020equivariant} detailed benchmarking compared to discrete NF layers and gradient flows has been performed on a similar problem showing the superiority of the equivariant continuous NF for particle systems over discrete ones.}  Continuous NFs transform the prior distribution into the target distribution by learning a time- and space-dependent vector field, $v(x_{v}(t),t)$, $t\in [0,1]$, which can be interpreted as force field,
\begin{equation}\label{eq:deq}
    \frac{\text{d}}{\text{d} t} x_v(t)= v (x_{v}(t),t), \quad x_{v}(0) = x_0 \text{ drawn from } \rho_P(x).
\end{equation}
We define the invertible transformation $F,$ as $x_v(t=1) \equiv F x_0$\footnote{Usually this transformation is defined as $T$, see Ref.~\cite{kohler2020equivariant}, which we avoid due to the importance of the temperature $T$ in the present study.}.
Importantly for the calculation of the loss function for training, we can evaluate the transformation of the prior probability distribution~\cite{kohler2020equivariant},
\begin{equation}\label{eq:density}
  	\log {\rho_F}(Fx_0) = \log \rho_P(x_0) - \int_{0}^{1} \text{d}t \text{ div } v(x_v(t),t).
\end{equation}
For a given transformation $F$, the NF thus produces a ``push-forward'' probability distribution ${\rho_F}(x)$ given by Eq.~(\ref{eq:density}), which is different from the target $\rho_*(x)$ if the transformation is not perfect. Similarly, ${\rho_{\bar{F}}}(x)$ emerges from the inverse transformation $\bar{F}$ when applied on the true distribution $\rho_*(x).$ 

The model used for the force field is a sum of pairwise potentials which depend on the distance between particle pairs~\cite{kohler2020equivariant},
\begin{align}
v(x(t),t) &= \nabla_x \Phi(x(t),t),\\
\Phi(x(t),t) &= \sum_{ij} \tilde{\Phi}(d_{ij}(t),t) ,
\end{align}
with $d_{ij}(t) = |x_i(t) - x_j(t)|$. The learnable weights $\{w\}$ of the NF are the parameters of the potential field $\tilde{\Phi}(d,t)$, which is parameterized using Gaussian radial basis functions in both distance $d$ and time $t$. \gj{The calculation of the divergence terms is numerically exact and stable,} as detailed in Ref.~\cite{kohler2020equivariant}.  Our implementation is based on the public code \href{https://github.com/noegroup/bgflow}{bgflow} provided by the authors of this publication. In the following, our goal is to transform an easy-to-sample high-temperature distribution, $\rho_P$, at inverse temperature $\beta_P$ into a low-temperature target distribution. 

\subsection{Loss function and training}
\label{sec:loss}

Normalizing flows can be trained using the Kuhlback-Leibler divergence as minimizable loss function $L$, which quantifies the similarity between the target distributions and the transformed NF distributions, $L= \alpha D_\text{KL}( {\rho_F} ||  \rho_* ) + (1-\alpha) D_\text{KL}( \rho_*|| \rho_{\bar{F}} ).$ We differentiate in $L$ between two different training contributions. The first term~\cite{kohler2020equivariant}, 
  		\begin{equation}
  	D_\text{KL}({\rho}_F ||  \rho_* ) =  \int_\Omega  \left[ \beta_*  U({x}) + \log {{\rho}_F}({x}) \right ] {{\rho_F}}({x}) dx,
  	\end{equation}
is based on having provided a set of high-temperature configurations, $ \{x_0\} $, which are transformed using $F,$ ${x} = F x_0.$ A loss based on $D_\text{KL}({\rho}_F ||  \rho_* )$ is called ``variational'' or ``energy-based'' training. This equation can be discretized as a sum over individual configurations $i$,
\begin{equation}\label{eq:energy-based}
  	  	D_\text{KL}({\rho}_F ||  \rho_* ) =  \sum_i  \left[ \beta_*  U({x}^i) - \int_{0}^{1} \text{d}t \text{ div } v(x_v^i(t),t) \right ],
\end{equation}
where we have dropped $\beta_P U(x_0^i) $ from Eq.~(\ref{eq:density}) since it is a constant that does not influence the loss function. 
   
It has been found in many studies that the training can be improved using a small set of low temperature configurations, $\{ x_*\}$ sampled from $\rho_*(x).$ The second term~\cite{kohler2020equivariant},
  	  		\begin{equation}
	D_\text{KL}( \rho_*|| \rho_{\bar{F}} ) =  \int_\Omega \left[- \beta_*  U( x_* ) - \log {\rho}_{\bar{F}}(x_*) \right ] \rho_*(x_*) dx_*
  	\end{equation}
quantifies the similarity between the transformed low-temperature configurations, $x = \bar{F}x_*$, and the prior distribution. This second contribution is also known as ``maximum likelihood'' training. We also discretize this equation to use it for training the NF.,
\begin{equation}
  	D_\text{KL}( \rho_*|| \rho_{\bar{F}} ) =  \sum_i \left[\beta_P U(\bar{F}x_*)  - \int_{0}^{1} \text{d}t \text{ div } v(x_v^i(t),t) \right].
\end{equation}
\gj{We have systematically analyzed the optimal value for $\alpha,$ as discussed in detail in App.~\ref{app:details}. We find that $\alpha=0.5$ leads to the most stable training procedure and the best final result (see Fig.~\ref{fig:alpha}). In particular, this figure also highlights the importance of including the maximum-likelihood training. } We also tested protocols in which $\alpha$ changes during the training procedure as suggested in Ref.~\cite{kohler2020equivariant}, but this did not affect the results qualitatively. 

Using a set of high- and low temperature configurations, $ \{x_0\} $ and $\{ x_*\}$, enables the optimization of the learnable parameters $\{w\}$, which quantify the strength of the force field $v(x(t),t)$. This step is achieved using the above-defined loss function and stochastic gradient decent. The errors are backpropagated using automatic gradient differentiation of the discretized NF in Eq.~(\ref{eq:deq}) implemented by PyTorch, which replaces the typical backpropagation known from artificial neural networks~\cite{kohler2020equivariant}. 

In our case, however, we face the problem that we do not have access to any low temperature configurations, because generating them is the whole purpose of the NF. We bypass this contradiction by using an iterative procedure. We use reweighting to generate some approximate low-temperature configurations. Using these configurations, we train a NF in the first iteration and apply it to create an improved set of low-temperature samples. In a second iteration we then utilize this improved set for the training of a second NF, which finally produces the low-temperature configurations that are analysed below. We can iteratively improve the performance of the NF with this iterative procedure but using more than two iterations did not lead to significant changes. In fact, we have tested that using low-temperature configurations prepared with SMC for training only marginally improved the performance of the NF, which thus confirms the efficiency of the proposed iterative procedure. \gj{In fact, our approach is similar in spirit to Ref.~\cite{midgley2023flow} which also avoids the usage of target configurations by applying annealed importance sampling.}

There are many hyperparameters that can be tuned to parameterize the NF and optimize the training procedure, including the number, location and time-discretization of the radial basis functions,  training parameters and batch sizes. However, we found that the results are not very sensitive to the explored choices of these parameters. See Appendix~\ref{app:details} for more details.

\subsection{Unbiasing the NF distribution}

After training the NF, we obtain a transformation $F$, which is used to transform all available high-temperature configurations, $x^i = F x_0^i$. Because the mapping performed by the NF is only approximate, the resulting set of configurations are biased, i.e. they do not exactly sample the Boltzmann distribution at temperature $T$. This can be corrected by performing an unbiasing step. We calculate the statistical weight of each transformed configuration as   
\begin{equation}
  W_i = \exp \left[ \beta_P U(x_0^i) - \beta_* U(x^i) + \int_{0}^{1} \text{d}t \text{ div } v(x_v^i(t),t) \right].
  \label{eq:RWNF}
\end{equation}
Similar to RW, we then use the weights $W_i$ to create a set of low temperature configurations that can sample the Boltzmann distribution. The NF therefore not only provides transformed configurations, but also their statistical weights $W_i$, which
describe the effective weights of each configuration $i$ at the low temperature $\beta_*.$ In direct RW, only the second step is performed, and thus NF has the potential to provide a large improvement over RW by first transforming the original configurations, resulting in larger statistical weights in Eq.~(\ref{eq:RWNF}). 
 
\subsection{Results: Sampling efficiency of NF}

\begin{figure}
    \centering
    \includegraphics[width=\linewidth]{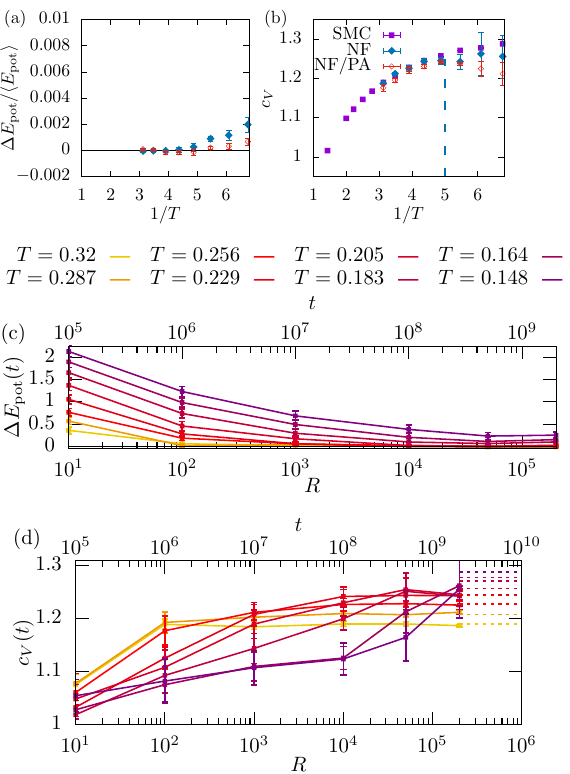}
    \caption{{\bf Benchmarking machine-learned normalizing flows (NF).} Description is the same as for Fig.~\ref{fig:PA}. For long-time averages in (a, b) we also show results for the combination of population annealing and normalizing flow (NF/PA). The vertical dashed line in (b) represents $T_{\rm NF}=0.2$ below which NF sampling fails.}
    \label{fig:NF}
\end{figure}

We now show the performance of the NF in Fig.~\ref{fig:NF} using the same metrics introduced for the other sampling methods, comparing long-time averages for the energy and the specific heat to SMC results. Regarding efficiency and timescales, we can analyze NF in much the same way as we did for PA in Sec.~\ref{sec:PA}. In particular, NF inherits the computational time $t = R \times 10^4 \tau$ of PA since it uses the same initial samples. Just as for PA, the sampling part of NF is computationally significantly more expensive than the subsequent transformation and unbiasing steps. 

We first compare NF to conventional MD results. From Fig.~\ref{fig:NF}(a,b), we conclude that the results for NF are much closer to the SMC groundtruth than what is achieved by MD simulations (see Fig.~\ref{fig:MD}). From Fig.~\ref{fig:NF}(b) we conclude that NF produces an equilibrium ensemble down to $T_\text{NF}=0.2$, which is significantly smaller than $T_\text{MD}=0.3$. Thus, the NF generative modeling approach is indeed an enhanced sampling method, in the sense that it works better than the physical dynamics in sampling low temperature configurations of the glassy system under study. Given published results regarding generative models for atomistic~\cite{coretti2022learning} or glass~\cite{ciarella2023generative} models, this is an interesting result. 

It is interesting to compare NF also with the direct RW approach studied in Sec.~\ref{sec:PA} as both methods use the same high-temperature configurations to predict low-temperature properties. The key step distinguishing the two methods is the NF transformation in Eq.~(\ref{eq:deq}) itself. The fact that NF performs much better than RW implies that the NF is able to efficiently transform the high-temperature configurations so that the transformed configurations are much closer to equilibrium than the original ones.  

As for PA, we can follow the approach to equilibrium of the potential energy, see Fig.~\ref{fig:NF}(c), and the specific heat, see Fig.~\ref{fig:NF}(d) when the size of the initial population $R$ is increased, which can be translated into timescale. These data allow us to define and measure a growing timescale for equilibration and sampling which becomes longer than the simulated time for $T < T_{\rm NF}$.  

Since NF outperforms MD simulations, it is pertinent to compare its performances with known enhanced techniques, which justifies our efforts to carefully benchmark various methods in Sec.~\ref{sec:benchmark}. Broadly speaking we find that all techniques (PA, PT, NF) perform nearly similarly, with NF and PA being slightly better than PT with the rough hierarchy, $T_\text{PA} \lesssim T_\text{NF} < T_\text{PT}.$ This detailed comparison and ranking of several techniques is one of the main results of this work: it provides evidences of the usefulness of NF for the difficult sampling problem of finite dimensional glassy systems. 

Given the success of NF over direct RW, it is tempting to combine the NF method with the successful PA approach in Sec.~\ref{sec:PA}, in order to possibly improve the performance of both these methods. In this combined approach, we use the global framework of PA, but we replace the second step in the PA algorithm (where copies are created from weights $W_i$ calculated using the Boltzmann distribution) by the usage of a trained NF to transform the configuration and calculate the new weights $W_i.$ We refer to this mixed method as ``NF/PA''.

The results shown in Fig.~\ref{fig:NF} are however disappointing. Although they are slightly better than NF, showing that multiple small steps are better handled than a large one, they are not better than PA. This is surprising, since NF clearly performs better than RW in the one-shot annealing procedure and PA is based on consecutive RW steps. Our interpretation is that for very small temperature steps, RW becomes actually superior to NF, presumably because it uses the exact expression of the Boltzmann distribution. 

\subsection{Analysis of the effective sample size}

Different from population annealing and parallel tempering, NFs have the capacity to produce low-temperature configurations in one shot, without the introduction of a large number of intermediate temperature steps. 

One practical consequence of such one-shot annealing is the possibility to define an interpretable effective sample size based on Kish's formula~\cite{kish1965survey},
\begin{equation}
\label{eq:kish}
    R_\text{eff} = \frac{ \left( \sum_{i=1}^R W_i \right)^2}{\sum_{i=1}^R W^2_i },
\end{equation}
using the statistical weights $W_i$ introduced in Eq.~(\ref{eq:RWNF}). The effective sample size describes roughly how many independent configurations have been produced during the annealing procedure.  The physical idea behind Eq.~(\ref{eq:kish}) is clear: the effective sample size $R_{\rm eff} \approx R$ if all weights $W_i$ are comparable, whereas $R_{\rm eff} \ll R$ when a few samples have a much larger weight than all others, indicating poor sampling. 

\begin{figure}
    \centering
    \includegraphics[scale=0.88]{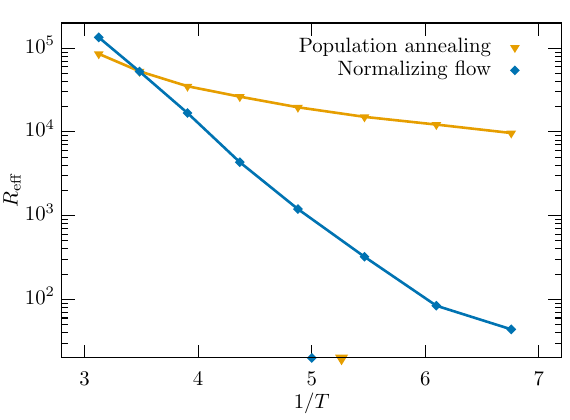}
    \caption{Effective sample size $R_\text{eff}$, as defined in Eq.~(\ref{eq:kish}), for the PA and NF results shown in Fig.~\ref{fig:PA} and Fig.~\ref{fig:NF}, respectively. The corresponding symbols on the x-axis mark the temperatures at which the sampling of each given method starts to fail.}    
    \label{fig:eff_samplesize}
\end{figure}

Starting from an initial set of $R=2 \times 10^5$ samples we observe for NFs an exponential decay of the effective sample size with temperature $T$ in Fig.~\ref{fig:eff_samplesize}. At the temperature $T_\text{NF}=0.2,$ identified before as the temperature at which NF sampling starts to fail, the effective sample size is $ 10^2 < R_\text{eff} < 10^3.$ This order of magnitude is consistent with our empirical findings that at least 100 independent samples are required for a proper representation of the equilibrium ensemble at temperature $T$. This analysis suggests that the effective sample size $R_\text{eff}$ can be used as an independent and easy tool to check for equilibration when using NF as an enhanced sampling technique.

In Fig.~\ref{fig:eff_samplesize}, we show the evolution of $R_\text{eff}$ evaluated during the gradual annealing employed for PA in Sec.~\ref{sec:PA}. We observe a much slower decrease when temperature is reduced, with an effective size that remains quite large, $R_{\rm eff} \sim 10^4$ when crossing the temperature $T_{\rm PA}$, indicating that the sample size is a poor indicator of adequate sampling in that case. This presumably results from the resampling of the algorithm whereby the samples that eventually dominate the low temperature behavior are replicated more often than others, which introduces strong correlations in the population. These correlations make the use of Kish formula inefficient. Other independent measures have been suggested for PA to test equilibration, but they require more involved analysis~\cite{amey2021measuring}.

\section{Discussion: What is the most efficient sampler?}

\label{sec:discussion}

\subsection{Quantitative comparison between techniques}

\label{sec:quantitative}

For each technique, we have obtained a temperature below which the assigned sampling task, i.e. measuring the specific heat, starts to fail. This allowed us to rank the various techniques. For the particular glass model studied here, we find that SMC is by far the best technique, with $T_{\rm SMC} = 0.12$. Then come the three enhanced algorithms, 
PA, NF and PT with $T_{\rm PA}=0.19$, $T_{\rm NF}=0.2$, $T_{\rm PT}=0.23$, which all perform much better than conventional MD with $T_{\rm MD}=0.3$. For comparison, we recall that the mode-coupling crossover is near $T=0.3$ and the experimental glass transition temperature near $T=0.15$~\cite{jung2022predicting}.

This ranking does not easily translate into an actual computational speedup, or efficiency gain, which may depend on the temperature. For each algorithm and each temperature, we showed that the approach to equilibrium of the energy or the convergence timescale for the specific heat can both be recorded to assign a representative sampling timescale. In practice, we use the former and define a timescale $\tau_c$ as $\langle \Delta E_\text{pot}(\tau_c) \rangle_{\rm neq} = 0.5$. \gj{As a rule of thumb, a smaller $\tau_c$ implies a smaller computational cost and thus improved performance of the technique.}

\begin{figure}
    \centering
    \includegraphics[scale=0.9]{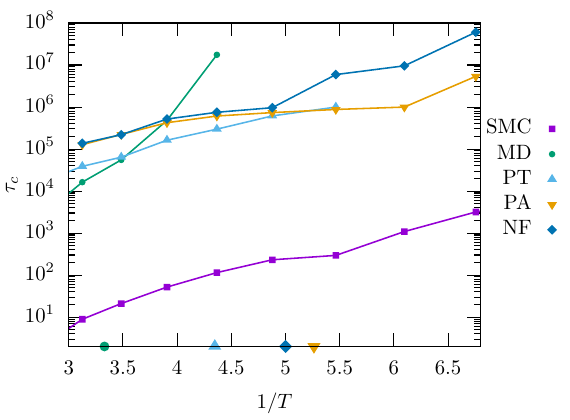}
    \caption{{\bf Temperature evolution of the efficiency timescale for all algorithms.} In practice, $\tau_c$ quantifies the approach to equilibrium of the potential energy. We compare swap Monte Carlo (SMC), molecular dynamics (MD), parallel tempering (PT), population annealing (PA), and normalizing flows (NF). The corresponding symbols on the x-axis mark the temperatures at which the sampling of each given method starts to fail. }
    \label{fig:timescale}
\end{figure}

We collect the results for the evolution of $\tau_c$ for all algorithms in Fig.~\ref{fig:timescale}. This provides a more detailed comparison between algorithms. Starting with very high temperatures, we observe in Fig.~\ref{fig:timescale} that MD is more efficient than the three enhanced sampling techniques, PT, PA and NF. In Sec.~\ref{sec:PT} we explained this finding for PT by the coupling between low and high temperatures through the temperature swap exchanges. The explanation is different for PA and NF which are less efficient due to the quite coarse sampling performed at high temperatures with a time $10^4 \tau$ between each stored configurations, see Sec.~\ref{sec:PA}. This time scale was the best compromise we found empirically between the time invested into the annealing and efficient sampling for the lowest temperatures. This could clearly be reduced if the focus was on higher temperatures. Given that MD is very efficient in this regime, this is not a crucial endeavour. 

When temperature decreases, Fig.~\ref{fig:timescale} shows that the MD timescale increases more rapidly than any other technique, and MD sampling is therefore the first to fail. The relaxation times of the three enhanced sampling methods, PT, PA and NF seem to roughly follow the same temperature dependence, with minor differences between them. Their behavior appears to be approximately Arrhenius, but the apparent energy barrier is much smaller than for MD. Notice that for PT the timescale $\tau_c$ does not take into account the fact that $n$ replicas need to be simulated in parallel. In the same vein, we note that the computational time for PA and NF is mostly due to the preparation of a large population of independent configurations at relatively high temperatures. This task can trivially be parallelized by running a large number of independent simulations, thereby making PA and NF potentially much more efficient than PT where no additional parallelization can be implemented.  

Interestingly, SMC seems to follow the same Arrhenius dependence of the three enhances methods, at least in this temperature regime, but with a prefactor that is considerably smaller by about four orders of magnitude. This large difference quantitatively explains why SMC is the most efficient sampling technique for this system. 

Despite the success of SMC, it is encouraging that NF can truly compete with state-of-the-art sampling techniques such as PT and PA, with a significant speedup over MD dynamics. At the lowest temperature where NF still operates, $T_{\rm NF}=0.2$, the speedup over MD dynamics is about four orders of magnitude in relaxation time.  

\subsection{Perspectives}

In this work we compared state-of-the-art enhanced sampling techniques for equilibrating supercooled liquids with a new method based on the machine learning technique using normalizing flows. Our results demonstrate the potentiality of ML methods to equilibrate model supercooled liquids at low temperature. In fact, the NF method applied to small systems at very low temperatures has a performance comparable to the sampling methods developed for complex systems, such as parallel tempering and population annealing. This very good result is obtained despite the fact that NF does not introduce a large set of replica (as in PT) or intermediate annealing temperatures (as in PA) and directly targets low temperatures in one shot. This positive conclusion suggests that all important modes of the low temperature states are already present, although affected by thermal fluctuations, in the high-temperature regime. However, NF are, like PT and PA, suboptimal with respect to the swap Monte Carlo technique.

We have focused on small systems with $N=43$ at very low temperatures in $d=2$. As demonstrated, this provides a challenging setting for all sampling methods for an atomistic model with realistic interactions. Applying the sampling methods to larger system sizes introduces new challenges for all of them. SMC and MD methods do not suffer too much with larger $N$, since the computational time increases linearly with $N$ while their performances do not degrade. The situation is different for PT, PA and NF, for different reasons. We provide in Appendix \ref{app:large} results with a system that is four times larger with $N=172$, showing poorer performances. Addressing the challenge of scaling these algorithms to larger system sizes should be the focus of a dedicated future work. In fact, even a very accurate NF method would eventually lead to an increasing level of rejection in the reweighting step for large sizes, as the statistical weight should scale as $\exp(-cN)$ with $c$ a finite constant, which account for a small difference between the generated distribution and the Boltzmann target. This generic argument does not take into account the complex nature of the glassy configuration space, which may very well lead to additional sampling issues at larger system sizes. 

Still, the observed performance of NF should encourage further work towards the development of improved techniques. For instance, it would be interesting to study more complex parametrization of the flow than the one we used. Possible candidates are: equivariant coupling flows~\cite{midgley2023se3}, which combine the efficiency of coupling flows while maintaining equivariance, equivariant flow matching~\cite{lipman2023flow,klein2023equivariant}, which uses alternative loss functions for training~\cite{felardos2023designing}, \gj{annealed flow transport Monte Carlo \cite{pmlr-v139-arbel21a}}, or approaches based on diffusion models~\cite{xu2022geodiff,zheng2023predicting,shu2023diffusion}. \gj{Additionally, it might be possible to combine NF layers with intermittent periods of SMC dynamics to create a stochastic normalizing flow as in Ref.~\cite{NEURIPS2020_41d80bfc}.}

The benchmarks outlined in this manuscript aim to accelerate and simplify the development of such sophisticated machine learning methods for sampling of complex systems. It is anticipated that any enhancement will manifest directly in the resulting relaxation time. Similar benchmarking for other complex system would be very valuable. We therefore believe that this manuscript marks an important step on the quest of finding methods that outperform traditional enhanced sampling techniques and, potentially, even the swap Monte Carlo technique.

\acknowledgments

We thank the No{\'e} group for publicly providing their library \href{https://github.com/noegroup/bgflow}{bgflow}, and S. Ciarella, M. Gabri\'e, and F. Zamponi for discussions . This work was supported by a grant from the Simons Foundation (\#454933, L. Berthier, \#454935 G. Biroli).

\appendix 

\section{Details on the NF method}

\label{app:details}

We provide more information on the hyperparameters used in the NF method. Our implementation is based on the bgflow library~\cite{noe2019boltzmann,kohler2020equivariant} (\href{https://github.com/noegroup/bgflow}{https://github.com/noegroup/bgflow}), which was extended to include periodic boundary conditions and multiple particle types. Thus any detail provided in Ref.~\cite{kohler2020equivariant} similarly applies to the present manuscript.

\subsection{Hyperparameter}

Most notably, we discretize the differential equation flow in Eq.~(\ref{eq:deq}) using just $N_t=1$ (first iteration) or $N_t=3$ timesteps (second iteration) in a multi-step fourth order Runge-Kutta scheme \cite{kohler2020equivariant}. This is a strong reduction of the complexity of the normalizing flows, but we have empirically found that larger values of $N_t$ do not improve the results. A consequence of this choice is that the transformation of the configurations amounts to quite small displacements $\Delta x \ll \sigma $ within the particle cages, rather than large-scale rearrangements. We have tried intensively to learn more general models, starting from $T\rightarrow\infty$ (uniformly distributed particles), but none of these models was able to propose acceptable configurations for low temperatures and reach accuracies comparable to the results presented in the main text.

We include 80 independent Gaussian radial basis functions centered non-uniformly at distances $d$ in the range $0.65 \leq d \leq 2.8.$ In total this adds up to 966 learnable parameters (1938 for $N_t=3$).  The gradient decent is based on an Adam optimizer with accuracy $10^{-4}.$ For the first iteration we train on 512 different structures using only one epoch, the second iteration uses four epochs and 4096 different structures. The batch size is always 64 structures.

\gj{\subsection{Ablation study for mixing parameter $\alpha$}}

\begin{figure}
    \centering
    \includegraphics[scale=0.9]{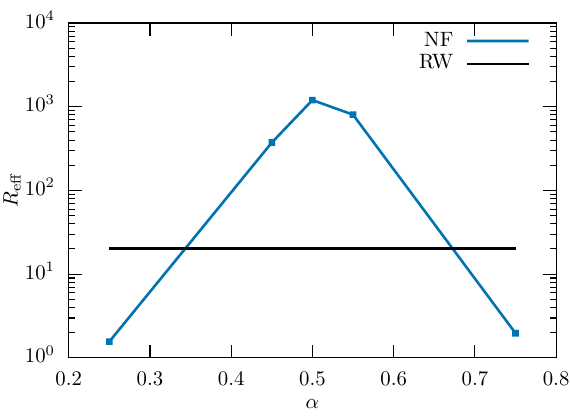}
    \caption{\gj{Effective sample size $R_\text{eff}$, as defined in Eq.~(\ref{eq:kish}), for different values of $\alpha$ at $T=0.205.$ The result for normalizing flows (NF) is compared to reweighing (RW), see Sec.~\ref{sec:PA}.}}
    \label{fig:alpha}
\end{figure}

\gj{We have introduced in Sec.~\ref{sec:loss} the parameter $\alpha$ in the loss function which interpolates continuously between energy-based training ($\alpha=1$) and maximum-likelihood training ($\alpha=0$). Which is the best choice for $\alpha$?}

\gj{To answer this question we have performed different training procedures for different values of $\alpha$. We report in Fig.~\ref{fig:alpha} the effective sample size $R_\text{eff}$ which we have identified in the main manuscript as an important factor to quantify the performance of the NF. The figure highlights a maximum near $\alpha \approx 0.5$, which is our final choice. For all other choices, the effective sample size is significantly lower.  }

\gj{In particular, Fig.~\ref{fig:alpha} rules out the possibility to perform pure energy-based training ($\alpha=1$) which would avoid the iterative procedure of finding low temperature training configurations described in Sec.~\ref{sec:loss}. In fact, we find that the problem with $\alpha=1$ is not mode-collapse as in other studies in the field of computer vision. For example, we have attempted the $\beta-$NF approach in which the entropy term (i.e., the second term in Eq.~(\ref{eq:energy-based})) is scaled by a factor $\beta > 1$  \cite{Sun_Bouman_2021,higgins2017betavae} and we did not find any improvement. Instead the only solution we found to increase $R_\text{eff}$ for $\alpha \neq 0.5$ is early stopping, which hints to some instabilities in the learning. Nevertheless, even with early stopping $R_\text{eff}(\alpha)$ never reaches the the value $R_\text{eff}(\alpha=0.5).$}

\gj{In conclusion, this analysis shows that $\alpha=0.5$ is the optimal choice for the mixing parameter.}\\

\section{Additional criteria for equilibration}

\label{app:equilibrium}

In the main text we state that the best way to validate sampling is by verifying whether $c_V(t)$ attains a plateau. During our research, we have tested several different possibilities, which we briefly describe. 

\subsection{Fluctuation-dissipation theorem for specific heat}

\begin{figure}
    \centering
    \includegraphics[scale=0.9]{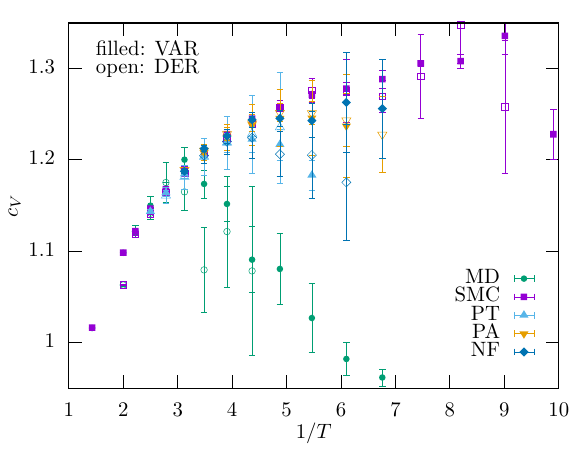}
    \caption{Specific heat $c_V$ calculated for various sampling techniques using two different definitions, $c^\text{VAR}_V$ from fluctuations and $c^\text{DER}_V$ from derivative. In equilibrium both are related by a fluctuation-dissipation theorem. The data indicate however that when the system falls out of equilibrium the fluctuation-dissipation theorem remains valid and both expressions similarly depart from equilibrium.}
    \label{fig:spec_heat}
\end{figure}

A popular way to validate equilibrium sampling is by calculating the specific heat using two different formula. The first one is used throughout this work and corresponds to the variance of fluctuations in potential energy, $c_V^\text{VAR} = (\langle E_\text{pot}^2 \rangle - \langle E_\text{pot} \rangle^2)/NT^2.$ A second definition is based on the temperature derivative of the average potential energy, $c_V^\text{DER} = N^{-1} \partial \langle E_\text{pot}(T) \rangle / \partial T . $ For equilibrium samples, these two definitions yield the identical result by virtue of the fluctuation-dissipation theorem. Any difference between these two quantities can therefore reveal a departure from equilibrium.

The results in Fig.~\ref{fig:spec_heat} show that this indicator does not clearly signal departure from equilibrium. In fact, it seems that when a given sampling technique departs from the SMC solution both definitions of the specific heat depart similarly at the same temperature, but remain consistent with each other within the error bar. A slightly better indicator of departure from equilibrium is the notable increase of the error bars, which indicate increasing correlations between configurations, indirectly revealing lack of ergodicity. It is however difficult to transform this observation into a clear-cut criterion for equilibration. 

\subsection{Probability distribution of potential energy}

We have analyzed in detail the average potential energy, $\langle E_\text{pot} \rangle$ and its variance in the form of the specific heat, $c_V.$ Here we investigate whether the full probability distribution of potential energies gives additional information, in particular on whether equilibrium sampling has been achieved.

\begin{figure}
    \centering
    \includegraphics[scale=0.9]{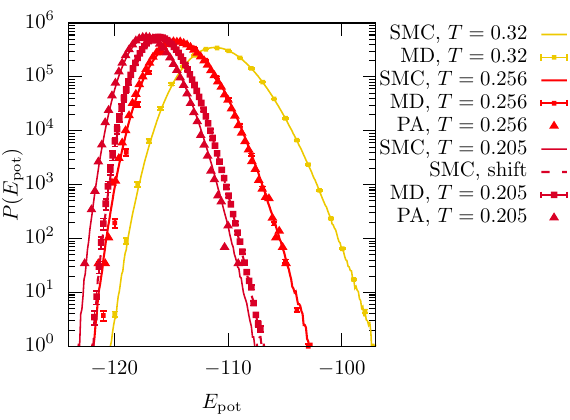}
    \caption{Probability distribution function of the potential energy, $P(E_\text{pot},T)$,  calculated for various sampling techniques at different temperatures $T$.}
    \label{fig:pE}
\end{figure}

We observe that histograms do not yield much more information compared to the first two moments, see Fig.~\ref{fig:pE}. At $T=0.256$ where equilibrium sampling already fails for MD, the energy histograms remain quite close, with small deviations only visible in the left tail at low energy values. At $T=0.205$, the MD dynamics are completely out-of-equilibrium which can be observed by a clear shift compared to the SMC result. However, by rescaling the first and second moment of the SMC distribution (dashed line) we observe nearly perfect overlap with the MD results (blue squares). 

We further exploit these data and evaluate the density of state $ G(E_\text{pot}) \propto P(E_\text{pot},T) \exp(\beta E_\text{pot})$. The interest of the density of state is that it is a temperature independent quantity which is only accurately obtained if proper equilibrium sampling of energy fluctuations is performed. As such it has been used as a tool to assess the degree of equilibration~\cite{PhysRevE.61.5473}. 

\begin{figure}
    \centering
    \includegraphics[scale=0.9]{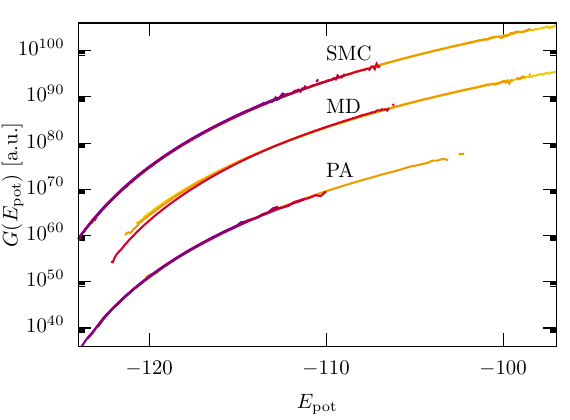}
    \caption{Density of states, $G(E_\text{pot})$,  calculated for various sampling techniques. For each technique, estimates of the $G(E_{\rm pot})$ obtained at different temperatures are stitched together to form a mastercurve. Each mastercurve is vertically shifted, for clarity.}
    \label{fig:GE}
\end{figure}

Our results are shown in Fig.~\ref{fig:GE}. Since the density of states is only known up to a prefactor, each set of curves is arbitrarily shifted to maximize the overlap between estimates of the density of states stemming from different temperatures for a given algorithm. In addition, the result for each method is shifted independently for better visualization.

The excellent data collapse for the SMC data confirms that equilibrium sampling is achieved down to very low temperatures. The expected temperature-independent mastercurve is obtained when stitching together the data from $P(E_\text{pot},T)$ obtained at different temperatures. 

Interestingly, the MD data indeed reveals, that the ensemble falls out of equilibrium since no perfect overlap can be achieved. This shows that the low-energy tails of the energy distribution are not properly sampled, in a way that is perhaps clearer than in Fig.~\ref{fig:pE}. 

In contrast, even at $T=0.148$, the data extracted from PA sampling shows perfect overlap although we know that they do not perfectly represent the equilibrium ensemble, as identified above. The reason for the qualitative difference between MD and PA is two-fold: (i) MD falls out of equilibrium much more violently, in particular when investigating the potential energy, while in PA the differences are much smaller even when the system is out of equilibrium. (ii) The number of samples used to create the histograms is much smaller in PA, since each sample needs to be treated independently, making it impossible to maintain a huge set. The very subtle difference in $G(E_\text{pot})$ observed for MD is therefore nearly invisible for PA.  

We conclude that $ G(E_\text{pot})$ can detect non-equilibrium properties, but it requires significant departure from equilibrium and huge datasets. In other words, this is not a very sensitive test of equilibrium sampling.

\subsection{Radial distribution function}

There are also two different ways to calculate the radial distribution function, $g(r),$ in particle systems at thermal equilibrium. The first traditional approach is based on histograms, and measures the density profile around a tagged particle. The second is based on forces, as recently proposed in Ref.~\cite{rotenberg2020force}. The identity between both methods is based on the assumption that the system is in thermal equilibrium. Therefore, any difference between the two expressions can be taken as the sign that the system is not equilibrated, but this approach has not be tested before. 

\begin{figure}
    \centering
    \includegraphics[scale=0.9]{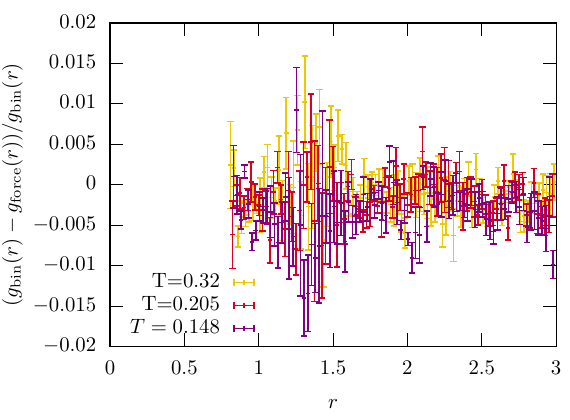}
    \caption{Relative difference between two expressions of the radial distribution function $g(r)$, calculated using either the traditional histogram method and the force method~\cite{rotenberg2020force}, for various temperatures. The small systematic differences are due to discretisation issues and do not depend on the degree of equilibration.}
    \label{fig:gr}
\end{figure}

Overall, we find that the relative difference between the two expressions for the pair correlation are extremely small, typically smaller than 1~\%, see Fig.~\ref{fig:gr}. A small systematic signal is observed when calculating the difference between both techniques. However, this signal depends on the discretization and binning of the histograms and is thus observable independently of the temperature. Apart from this signal, no systematic difference between the two computation methods can be observed. This method therefore cannot be used to detect non-equilibrium properties. 

This result is reminiscent of similar findings for the configurational temperature,  which is shown to decay instantaneously to the thermal temperature during equilibration~\cite{Dyre2021aging}. The relationship between the histogram and the force methods for $g(r)$ corresponds roughly to a space-dependent generalization of the global configurational temperature.

\section{Scaling with system size}

\label{app:large}

\begin{figure}[t]
    \centering
    \includegraphics[]{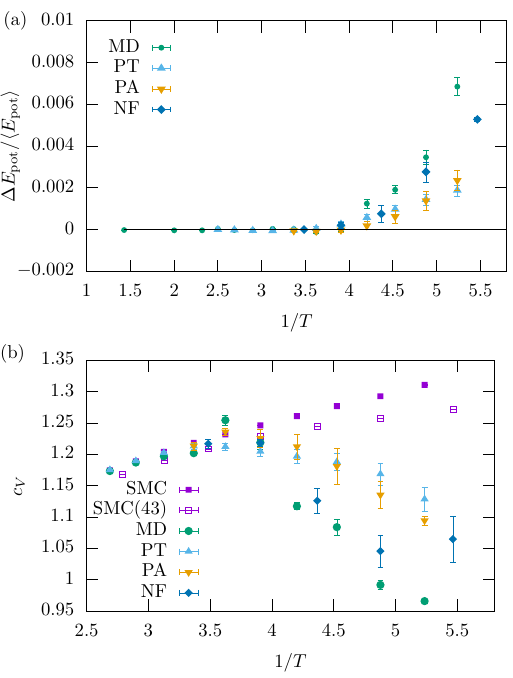}
    \caption{Scaling the results to a larger system size, $N=172$. (a) Difference in potential energy from the SMC result $\Delta E_\text{pot} = \langle E_\text{pot} \rangle - \langle E^\text{SMC}_\text{pot} \rangle $. (b) Specific heat $c_V$. The open symbols show the SMC results for $N=43$, for comparison.}
    \label{fig:172}
\end{figure}

In the main text, we concentrate on a small system size, $N=43$, and we only briefly mention how the results may change with system size. 

We repeated sampling with SMC, MD, PT, PA and NF for a larger system size, $N=172$. As in the main text, we then use the SMC results as a benchmark and report in Fig.~\ref{fig:172}(a) deviations of the average energy with respect to SMC. In   Fig.~\ref{fig:172}(b) we show results for the specific heat for the various sampling methods. 

Compared to $N=43$, the performance of the MD approach are essentially the same, with deviations appearing near $T_{\rm MD}=0.3$ in both quantities. However, we observe that the efficiency of the three enhanced sampling techniques (PT, PA, and NF) is significantly reduced in larger systems, as expected~\cite{falcioni1999biased,earl2005parallel}. In detail, we see that PT and PA now have a comparable performance, with a speedup compared to MD that is much less impressive than for $N=43$ particles. This strong decrease in performance for both techniques stems from the complexity of sampling multiple low-energy states in glassy systems.  

We also conclude that the normalizing flows suffer from the same reduction in performance with increasing system size. Therefore, our current implementation of NF does not get more efficient in larger systems compared to traditional enhanced sampling techniques such as PT and PA. Scaling the NF method to large sizes is clearly a challenging problem, which therefore deserves further attention in future work. 

\clearpage

\bibliography{library_local.bib}

\end{document}